# 3D Structural Phenotype of the Optic Nerve Head at the Intersection of Glaucoma and Myopia – A Key to Improving Glaucoma Diagnosis in Myopic Populations

Swati Sharma[1], Fabian A. Braeu[1,2,3], Thanadet Chuangsuwanich[1,4,5], Tin A. Tun[1,3], Quan V Hoang[1,3,4], Rachel Chong[1,3], Shamira Perera[1,3], Ching-Lin Ho[1,3], Rahat Husain[1,3], Martin L. Buist[6], Tin Aung[1,3,4], Michaël J. A. Girard[1,3,5,7,8]

1. Singapore Eye Research Institute, Singapore National Eye Centre, Singapore
2. Singapore-MIT Alliance for Research and Technology, Singapore
3. Duke-NUS Graduate Medical School, Singapore
4. Yong Loo Lin School of Medicine, National University of Singapore, Singapore
5. Department of Ophthalmology, Emory University School of Medicine, Emory University
6. Department of Biomedical Engineering, National University of Singapore, Singapore, Republic of Singapore
7. Department of Biomedical Engineering, Georgia Institute of Technology/Emory University, Atlanta, GA, USA
8. Emory Empathetic AI for Health Institute, Emory University, Atlanta, GA, USA

**Corresponding Author**:

Michaël J.A. Girard

Ophthalmic Engineering & Innovation Laboratory

Emory Eye Center, Emory School of Medicine

Emory Clinic Building B, 1365B Clifton Road, NE

Atlanta GA 30322

mgirard@ophthalmic.engineering

**Abbreviations:**

RNFL: Retinal nerve fiber layer

GCL+IPL: Ganglion cell layer with the inner plexiform layer

LC: Lamina cribrosa

ONH: Optic nerve head

IOP: Intraocular pressure

OCT: Optical coherence tomography

3D: Three-dimension

AUC: Area under the receiver operating characteristic curve

2D: Two-dimension

H: Healthy

HM: Highly myopic

G: Glaucoma

HMG: High myopia and glaucoma

MLP: Multilayer perceptron

PCA: Principal component analysis

PD: Principal directions

SCO: Scleral canal opening

NT: Nasal-temporal

MRW: Minimum rim width

BM: Bruch's membrane

BMO: Bruch's membrane opening

## Abstract:

**Purpose**:

To characterize the 3D structural phenotypes of the optic nerve head (ONH) in patients with glaucoma, high myopia, and concurrent high myopia and glaucoma, and to evaluate their variations across these conditions.

**Design**: Retrospective cross-sectional study.

**Participants:** A total of 685 optical coherence tomography (OCT) scans from 754 subjects of Singapore-Chinese ethnicity, including 256 healthy (H), 94 highly myopic (HM), 227 glaucomatous (G), and 108 highly myopic with glaucoma (HMG) cases

**Methods**: We segmented the retinal and connective tissue layers from OCT volumes and their boundary edges were converted into 3D point clouds. To classify the 3D point clouds into four ONH conditions, i.e., H, HM, G, and HMG, a specialized ensemble network was developed, consisting of an encoder to transform high-dimensional input data into a compressed latent vector, a decoder to reconstruct point clouds from the latent vector, and a classifier to categorize the point clouds into the four ONH conditions. Additionally, the network included an extension to reduce the latent vector to two dimensions for enhanced visualization.

**Main Outcome Measures**: Structural variation in the ONH in H, HM, G, and HMG conditions

**Results**: The classification network achieved high accuracy, distinguishing H, HM, G, and HMG classes with a micro-average AUC of 0.92 ± 0.03 on an independent test set. The decoder effectively reconstructed point clouds, achieving a Chamfer loss of 0.013 ± 0.002. Dimensionality reduction clustered ONHs into four distinct groups, revealing structural variations such as changes in retinal and connective tissue thickness, tilting and stretching of the disc and scleral canal opening, and alterations in optic cup morphology, including shallow or deep excavation, across the four conditions.

**Conclusions**

This study demonstrated that ONHs exhibit distinct structural signatures across H, HM, G, and HMG conditions. The findings further indicate that ONH morphology provides sufficient information for classification into distinct clusters, with principal components capturing unique structural patterns within each group.

# 1 Introduction:

Glaucoma is a chronic, progressive optic neuropathy marked by complex structural changes in the retina. The primary pathological event is the gradual deterioration of tissue, particularly the retinal nerve fiber layer (RNFL), the ganglion cell layer with the inner plexiform layer (GCL+IPL), and neuroretinal rim thinning. [1, 2] Other structural changes include alterations in lamina cribrosa (LC) depth and curvature, and peripapillary scleral bowing. [3, 4] These damages are irreversible, and when undetected at early stages, the loss of retinal ganglion cells (through apoptosis) can lead to permanent vision loss.

High myopia has been extensively reported as a significant risk factor for glaucoma with morphological changes in the optic nerve head (ONH). Epidemiological studies, including the Beijing Eye Study, [5] Blue Mountains Eye Study, [6] Barbados Eye Study, [7] and Malmö Eye Survey [8] have all documented a higher prevalence of glaucoma in eyes with high myopia and the correlation of ONH damage with the severity of myopia. These structural changes, frequently difficult to distinguish from glaucomatous damage, could contribute to the increased prevalence of glaucomatous optic neuropathy in high myopia.

Optical coherence tomography (OCT) has been a widely used imaging modality for diagnosing glaucoma due to its ability to capture the complex three-dimensional (3D) structure of the ONH. [9] However, the concurrent presence of both glaucomatous and high myopic conditions introduces structural variations that complicate OCT evaluation. [10-12] Zemborain et al. examined morphological parameters from OCT images to differentiate between glaucoma and high myopia, concluding that this differentiation largely depends on the clinician's expertise in interpreting the scans. [11] Similarly, Rezapour et al. used structural features, such as RNFL and GCL+IPL thickness, and minimum rim width (MRW), for classification between glaucoma and high myopia. [12] However, these studies focused on limited one-dimensional structural parameters that are clinically accepted. By using these types of parameters, the possibility of exploring new and novel biomarkers is overlooked, and it also fails to pay full attention to the intricate 3D morphology of the ONH.

In defining the structural phenotype of the glaucomatous ONH, Panda and colleagues employed an autoencoder-based deep learning model that compressed and decompressed input data to learn crucial features and simultaneously reconstructed and classified segmented OCT images into healthy and glaucomatous categories. [13] This approach enabled the identification of novel biomarkers from segmented OCT images, which not only demonstrated high diagnostic accuracy but also offered insights into structural differences between healthy and glaucomatous conditions. While the network effectively reconstructed and extracted key features from two-dimensional (2D) images, it was not designed to capture the 3D characteristics of the ONH from these 2D segmented slices.

A more comprehensive approach involves leveraging the full 3D OCT volume to train deep learning networks for classification tasks, though such networks often come with significant computational overhead. [14, 15] To address this, a novel method using 3D point cloud representations of the ONH has been introduced. [16, 17] This approach efficiently captures the 3D morphology of the ONH while reducing redundant OCT scan information. Such an approach has also allowed us to (1) diagnose glaucoma robustly, better than other gold-standard AI techniques, [18] (2) identify key neural and connective tissue landmarks (distributed in 3D) that contribute to such a diagnosis, [19] (3) emphasize the importance of biomechanics in glaucoma pathophysiology. [20]

This study aimed to integrate two of our recent AI advancements, i.e., an autoencoder for identifying key structural signatures of the ONH and the application of PointNet inspired networks for comprehensive 3D analysis of the ONH point cloud, to characterize the structural phenotype of the ONH at the intersection of glaucoma and myopia. Specifically, the objectives were: (1) to develop an autoencoder capable of extracting essential structural phenotypes of the ONH from 3D point clouds by compressing and reconstructing input data through an encoder-decoder framework; (2) to enhance the network's capability to classify eyes into four categories: healthy (H), highly myopic (HM), glaucomatous (G), or a combination of high myopia and glaucoma (HMG); and (3) to elucidate the key structural differences in the ONH associated with these conditions.

## 2 Methods

### 2.1 Patient recruitment and OCT imaging

A total of 754 subjects participated in a cross-sectional study conducted at the Glaucoma Clinics, Singapore National Eye Centre. Inclusion criteria involved subjects of Singapore-Chinese ethnicity aged over 50 years, with exclusion criteria, such as no prior intraocular/orbital/brain surgeries, no known history of medical strabismus, ocular trauma, ocular motor palsies, orbital/brain tumors, pathological myopia, clinically abnormal saccadic or pursuit eye movements, poor LC visibility in OCT (< 50% enface visibility), known carotid or peripheral vascular disease, or other abnormal ocular conditions. Subjects with corneal abnormalities and cataract that have the potential to preclude the quality of the scans were excluded from the study. Furthermore, subjects with pathological myopia were also excluded from the study. This resulted in a total of 685 OCT scans with 256 H, 94 HM, 227 G, and 108 HMG eyes.

The healthy eyes were defined as those with a refractive error between +2.75 and −5 diopters, an axial length of less than 25 mm, IOP below 21 mmHg, and no pathological changes in the ONH or retina. High myopia was characterized by an axial length exceeding 25 mm and a refractive error less than −5 diopters. HMG referred to eyes that satisfied the criteria for HM and received a glaucoma diagnosis from experts (authors RC and TAT). Glaucomatous eyes exhibited a loss of the neuroretinal rim, with a vertical cup-to-disc ratio greater than 0.7. For further details, readers are encouraged to consult our earlier publication. [21]

All subjects provided written informed consent, and the study followed the principles of the Declaration of Helsinki, approved by the SingHealth Centralized Institutional Review Board. Table 1 summarizes the subjects population.

OCT scanning was performed on each participant using a spectral-domain OCT system (Spectralis, Heidelberg Engineering, Germany). Scans were conducted in a dark room after dilating the pupils with 1.0% Tropicamide. The imaging protocol included a raster scan centered on the ONH, covering a rectangular area of 15 degrees X 10 degrees. Each OCT volume consisted of 97 B-scans, with each B-scan containing 384 A-scans and 496 pixels per A-scan. The average spacing between adjacent B-scans was 35.1 µm, while the axial and lateral resolutions within a B-scan averaged 3.87 µm and 11.5 µm, respectively. More details on the imaging methodology can be found in our earlier publications. [22, 23]

## 2.2 Point cloud creation from OCT volumes

A point cloud is an efficient representation for capturing complex 3D geometries in a sparse format. This approach retains the 3D morphology while significantly reducing computational costs. In order to construct a point cloud from an OCT volume, we first used the software Reflectivity (Abyss Processing Pte Ltd, Singapore) to segment the retinal and connective tissue layers of the ONH, i.e., retinal nerve fiber layer (RNFL), ganglion cell layer and inner plexiform layer (GCL+IPL), all other retinal layers (ORL), retinal pigment epithelium (RPE), choroid, the visible portion of the peripapillary sclera (encompassing the scleral flange), and the lamina cribrosa (LC). Reflectivity also registered all OCT volumes with respect to their Bruch's membrane opening (BMO) planes and BMO centers to facilitate all structural comparisons. Fig. 1a illustrates a raw OCT volume, Fig. 1b shows a representative B-scan from the volume, and Fig. 1c depicts the segmented tissue layers of the ONH.

Subsequently, the anterior and posterior boundaries of the RNFL, prelamina, ORL, sclera, and LC were identified and converted into a set of 3D points (see Fig. 1d and 1e). This process was repeated for all B-scans in the OCT volume to construct a 3D point cloud, providing a sparse representation of the complex 3D structure of the ONH (Fig. 1f). Finally, the point clouds were down sampled to 4096 points (see Supplementary material for more details on point cloud creation). Furthermore, each point in the cloud was assigned a marker indicating the tissue layer from which it was extracted, enabling layer visualization by assigning unique colors to different layers (see Fig. 1f).

## 2.3 An ensemble autoencoder for structural phenotype identification

An autoencoder is a special type of neural network designed to learn compact and efficient representations of high-dimensional data (such as 3D point clouds). [24] It consists of two components: an encoder, which compresses the input data into a smaller latent representation by capturing its most essential features, and a decoder, which reconstructs the original input from this compact representation as accurately as possible. This structure enables the autoencoder to reduce noise and extract meaningful patterns in the data.

*Network Architecture*:

Numerous deep learning architectures have been proposed in the literature to analyse complex 3D geometries represented as 3D point clouds. Qi et al. proposed a novel PointNet architecture designed to process 3D point cloud for tasks like classification and segmentation. [17] Groueix and co-authors constructed a specialized network, called AtlasNet, to reconstruct 3D point clouds using 2D parametric patches. [25] In this study, we developed a specialized ensemble autoencoder network using: (1) PointNet as the encoder; (2) AtlasNet as the decoder; and (3) a network with dense layers as the classifier (see Fig. 2). The encoder gradually compressed the input point cloud into a reduced vector (also known as latent vector) of dimension k, capturing its most essential features. Subsequently, the decoder reconstructed the original point cloud using the latent vector (see Supplementary material for more details on network architecture). The classifier branch also utilized the latent vector to predict the final classes (i.e., H, HM, G, and HMG).

*Latent vector dimension optimization*:

To analyze the impact of the latent vector size, we experimented with different dimensions, i.e., k = 128, 256, 512, and 1024. For each k value, the network was trained independently, and the outputs were compared to evaluate the best dimension for the latent vector.

*Dataset split*:

We split the dataset into training (70%), validation (15%), and test (15%) sets, respectively. During this split, we also ensured that duplicate point clouds or point cloud of same subject should not appear in multiple sets. Additionally, we balanced the training set across the different studies and classes to avoid bias towards a particular class or study.

*Loss function and hyperparameters*:

The ensemble network minimized a loss function that was composed of a weighted sum of two components: the point cloud reconstruction loss (Chamfer loss) from the decoder branch and the categorical cross-entropy loss from the classification branch. Chamfer loss measures the similarity between two point clouds by calculating how close each point in one cloud is to its nearest neighbor in the other, considering both point clouds. A smaller Chamfer loss indicates greater similarity, with a value of zero signifying a perfect match between the point clouds. Conversely, categorical cross-entropy loss evaluates how accurately a model's predicted probabilities align with the true categories in classification tasks. A high predicted probability for the correct category results in a small loss, while a low probability increases the loss. We assigned different weights to these loss functions and examined their effects on the final output.

The total loss was minimized using ADAM optimizer with a learning rate of 0.001. All three branches were trained simultaneously on a Nvidia A5000 GPU card for more than 1,000 epochs till the total loss function converged, and the weights from the epoch that had the lowest validation loss considered as the optimal weights for the network.

*Network performance evaluation*:

The network performance was evaluated using the area under the receiver operating characteristic curve (AUC) for the classification branch and the chamfer loss for the reconstruction branch. Furthermore, a five-fold cross validation was performed to check the robustness of the network.

**2.4 Describing the structural phenotype of the ONH across health and disease**

The latent vector encodes critical structural information about the point cloud that is essential for both point cloud reconstruction and classification. We posit that the decoder's ability to recreate a point cloud highly similar to the input point cloud using only the k features of the latent vector underscores the significance of these encoded features. Thus, our one of the objectives was to investigate how variations in the features of the latent vector correspond to specific structural changes in the ONH, such as alterations in RNFL thickness or disc size.

To facilitate this analysis, we incorporated a separate branch into the ensemble network to further reduce the latent vector of size k into a size of two using Principal Component Analysis (PCA). The two resulting dimensions, referred to as principal directions (PDs), enabled the visualization of 3D point clouds as 2D points in a scatter plot for improved interpretability (see Fig. 2). Notably, the points in the 2D scatter plot formed different clusters, and subsequently, the cluster boundaries along with their centroid positions were identified using the k-means clustering approach. This branch was excluded from the training process and was used exclusively for dimensionality reduction and visualization.

The PCA approach also provides an inverse transformation function, enabling the back computation of the k-dimensional latent vector from the two PDs. To explore structural differences in the ONH across different clusters, we selected a point from one cluster and incrementally altered the values of its PDs over twenty steps to transition it toward a different cluster. At each step, the k-dimensional latent vector was computed from the new PDs using the inverse PCA transformation, and the decoder reconstructed a point cloud from the updated latent vector. These reconstructed point clouds were then compared to the baseline to identify structural changes. This process facilitated the visualization of potential ONH structural differences among clusters, such as those observed when transitioning from the healthy cluster to the glaucoma cluster or other conditions.

Visualizing subtle structural changes in the 3D point cloud during transitions between clusters proved challenging, as scatter points in 3D space lacked sufficient clarity. To address this limitation, we focused on the set of points corresponding to the central B-scan of the OCT volume (Fig. 3). These points were isolated and fitted with univariate splines to generate smooth curves representing the layer boundaries. The spline curves (Fig. 3b) provided a clearer representation of layer boundaries and their transformations during transitions between clusters, enabling a more detailed visualization of the structural changes.

### 2.5 Diagnostic capabilities of individual nerve layers

The structural signature of individual retinal and connective tissue layers plays a critical role in diagnosis, as conditions like myopia and glaucoma may impact these layers differently. To evaluate the diagnostic potential of the morphology of each layer, we generated separate point clouds for each layer and independently trained the ensemble network on the point cloud data of each layer. The resulting classification accuracies were compared, providing insights into the diagnostic capabilities of each layer.

## 3 Results

### 3.1 Classification network performance

Our classification network demonstrated high accuracy in distinguishing among H, HM, G, and HMG classes. On an independent test dataset, the micro-average AUC during 5-fold cross-validation was 0.92 ± 0.03, with individual AUC values: 0.95 ± 0.01 for H, 0.91 ± 0.03 for HM, 0.92 ± 0.03 for G, and 0.89 ± 0.03 for HMG classes (Fig. 4a-b).

The classification accuracy using individual layers revealed that the RNFL layer was the most critical for distinguishing between classes. When using only the RNFL, the AUC during 5-fold cross-validation was 0.89 ± 0.02. Other tissue layers also demonstrated diagnostic potential, with AUCs of 0.87 ± 0.04 for GCL+IPL and ORL, 0.78 ± 0.02 for the choroid, and 0.85 ± 0.03 for the sclera and LC. Table 2 provides a summary of the AUCs obtained during 5-fold cross-validation.

### 3.2 Reconstruction network performance

The decoder was able to reconstruct the point clouds accurately with Chamfer loss of 0.013±0.002 on the test dataset. Fig. 4c shows an original point cloud from the test dataset and the corresponding reconstructed point cloud using the decoder, and Fig. 4d depicts the box plot for chamfer loss on test set. We observed that the quality of point cloud reconstruction changed with the size of the latent vector. The Chamfer loss for latent vector sizes of 128, 256, 512, and 1024 were 0.024±0.006, 0.019±0.004, 0.013±0.002, and 0.014±0.002, respectively. So, the latent vector dimension was set at 512 for all further simulations.

### 3.3 Clustering of point clouds

The dimensionality reduction branch reduced the latent vectors to two dimensions using PCA, enabling the visualization of point clouds as a 2D scatter plot (Fig. 5). The plot displayed four distinct clusters corresponding to the four classes: green dots for H, red dots for G, cyan dots for HM, and black dots for HMG (Fig. 5a). The k-means clustering approach successfully identified the boundaries of the clusters. The centroids of each cluster are highlighted in Fig. 5a using star markers.

### 3.4 Structural signatures in glaucoma and high myopia

Distinct structural differences were observed in the ONH in different clusters. For instance, to examine structural changes and illustrate how the healthy ONH structure differs from the highly myopic ONH qualitatively, we morphed a healthy ONH into a highly myopic ONH, thereby highlighting the key differences. For this purpose, a point near the centroid of the healthy cluster (depicted as a large white dot with a green edge in Fig. 5b) was selected and its coordinates (PD1 and PD2 values) were gradually adjusted over twenty steps, directing the point toward the centroids of other clusters (indicated by black arrows showing the direction of morphing). At each step, the decoder reconstructed a point cloud from the updated latent vector obtained through inverse PCA transformation of PD1 and PD2. The inset images illustrate the central B-scan of the reconstructed point cloud near the centroid of each cluster. Similarly, Fig. 6 presents central B-scan images captured during the morphing of ONH between different clusters.

*ONH Structural Transformations – From Healthy to Other Conditions*

Healthy ONHs were characterized by a thick RNFL, thick prelamina, small cup diameter, shallow cup, and thick LC (Figs. 6a–6c). Transitioning from the healthy to the glaucoma cluster (Fig. 6a) resulted in significant structural changes, including RNFL thinning, prelaminar thinning, an increased cup diameter, a deeper cup, and LC thinning. Transitioning from the healthy to the high myopic cluster (Fig. 6b) resulted in ONHs with prelaminar thickening, disc elongation with nasal-temporal (NT) tilting, scleral thinning, and scleral canal opening (SCO) tilting in the NT direction, along with LC thinning. Similarly, transitioning from the healthy to the HMG cluster (Fig. 6c) led to RNFL thinning, prelaminar thinning, LC thinning, scleral thinning with a tilted SCO, disc elongation with NT tilting, and an increased cup depth. Similarly, transitioning from the healthy to the HMG cluster (Fig. 6c) demonstrated distinct structural modifications, including RNFL thinning, prelaminar thinning, LC thinning, scleral thinning with a tilted SCO, an elongated disc with NT tilting, and a deep cup.

***ONH Structural Transformations – From Glaucoma to Other Conditions***

Figs. 6d and 6e present a sample ONH structure from the glaucoma cluster. After transitioning to the HM and HMG clusters, the resulting ONHs exhibited distinct structural differences. The HM ONH displayed an elongated disc with pronounced tilting and stretching in the NT direction, accompanied by scleral thinning, shallowing of the cup, and prelamina thickening (Fig. 6d).

The HMG ONH showed similar changes to those of the HM ONH but was differentiated by a deeper cup than the glaucomatous ONH and a thinner prelamina (Fig. 6e). In both cases, the most prominent features included RNFL and scleral thinning, as well as disc and SCO tilting.

***ONH Structural Transformations – High Myopia Developing Glaucoma***

Fig. 6f illustrates an ONH from the HM cluster, highlighting characteristic structural features of high myopic ONHs, including scleral flange stretching and tilting of the disc and SCO. After transforming to the HMG cluster, the ONH exhibited pronounced glaucomatous changes, such as RNFL thinning, prelamina thinning, and a deeper cup, alongside myopic changes, including scleral thinning and further SCO stretching and tilting.

**4 Discussion**

In this study, we utilized point clouds to sparsely represent the intricate morphology of the ONH and evaluated their effectiveness in distinguishing structural variations associated with glaucomatous and highly myopic conditions. The analysis was conducted in two stages. First, a deep learning autoencoder network was employed to compress and decompress 3D point clouds of the ONH, capturing critical structural signatures during the process. Second, these structural signatures were used to classify ONHs into clusters representing H, G, HM, and HMG conditions. Our network demonstrated very good accuracy in point cloud compression-decompression and classification. The findings further revealed that the morphology contains sufficient information to classify ONHs into distinct clusters, with principal components highlighting unique structural signatures within each cluster.

Our proposed model achieved high classification accuracy in differentiating H, G, HM, and HMG ONHs, with an AUC of 0.92 ± 0.03, matching or surpassing results reported in similar studies. [11-15] Unlike conventional OCT imaging methods that rely on 2D B-scans or standard RNFL measurements, our approach leverages the full 3D morphological structure of the ONH influenced by both neural and connective tissues. Prior studies, including those by Akashi et al. and Shoji et al., have demonstrated the potential of OCT imaging and AI-driven analysis in distinguishing glaucomatous from myopic ONH changes. [26,27] However, limited research has specifically focused on the simultaneous classification of H, G, HM, and HMG ONHs. Our findings provide strong evidence that ONHs can be effectively categorized into these distinct classes based solely on their 3D morphology, offering a novel avenue for improving glaucoma diagnosis in highly myopic eyes.

Our autoencoder network achieved high reconstruction accuracy, with a Chamfer loss of 0.013 ± 0.002. It effectively extracted critical latent features from 3D point clouds essential for reconstruction and successfully captured key structural differences among H, G, HM, and HMG ONHs. Furthermore, our network demonstrated the critical morphological differences that occur when an ONH transformed from one cluster to another. Similar structural variations have been reported in 2D segmented images of the ONH by Panda et al. [13] in the context of glaucoma. However, to the best of our knowledge, no prior studies have illustrated these structural transformations across multiple ONH conditions in a 3D framework.

Diagnosing glaucoma in highly myopic eyes presents significant challenges due to overlapping structural alterations. Our study identified key structural differences when comparing G and HM ONHs. Glaucomatous ONHs exhibited a thinner RNFL and prelamina, along with a deeper cup and larger cup diameter. Additionally, the LC and sclera were thicker compared to HM ONHs (Figs. 6a, 6b, 6d, and Fig. A1 in the Appendix). These findings are consistent with numerous studies that have reported similar glaucomatous changes, such as increased cup-to-disc ratio, optic disc cup enlargement, RNFL thinning, and LC alterations. [28-34] In contrast, HM ONHs displayed distinct characteristics, including a tilted or obliquely inserted optic disc in the NT direction, tilting of the SCO in the same direction, and disproportionately wide, large, and shallow optic cups. These findings align with previous research by Kim and Park, who observed a stretched and temporarily enlarged Bruch's membrane (BM) opening, tilted disc, and thinner LC in HM ONHs. [35] Additionally, progressive optic disc tilting with nasal shifting, shallow cupping, and SCO tilting have been widely reported as key features of HM ONHs. [36,37] Hu et al. and AttaAllah et al. also highlighted ONH changes in highly myopic eyes due to axial elongation, such as higher cup-to-disc ratios, tilted ONHs, LC deformation, and enlarged optic discs. [26,27,34,38] The structural changes identified in our study are consistent with these observations.

When comparing G and HMG ONHs, we observed several distinct structural differences. The HMG ONH exhibited a thinner RNFL and LC than the G ONH, along with a thinner sclera. Additionally, HMG ONHs displayed focal cupping, deep excavation within the optic cup, and significant thinning of the prelaminar tissue. The most distinguishing features between G and HMG ONHs were the tilted and elongated optic disc in the NT direction and the tilted SCO (Fig. 6e and Fig. A1 in the Appendix). Furthermore, scleral flange stretching was noted in HMG ONHs. These findings align with prior studies. Zhang et al. identified key differences between HMG and G ONHs, including an overhanging BM on the nasal side of the optic disc, its absence on the temporal side, significantly smaller MRW, larger SCO, thinner LC, and a higher frequency of LC defects. [39] Similarly, Tan et al. reported significant neuroretinal, RNFL, and scleral thinning in HMG ONHs [40].

We observed a few common features, such as tilted and elongated optic discs in the NT direction, along with SCO and scleral flange stretching in both HM and HMG ONHs. However, the degree of tilting in both the SCO and optic disc was notably more pronounced in the HMG cluster compared to the HM cluster. Key differentiating features of HMG ONHs included focal cupping, deep optic cup excavation, thinning of the prelaminar tissue, and a significantly smaller MRW. These findings are consistent with previous studies. Jiravarnsirikul et al. emphasized that scleral flange stretching in highly myopic eyes plays a crucial role in ONH morphological changes, particularly in the development and expansion of exposed neural canal regions. [41-42] Additionally, Jonas et al. examined the impact of parapapillary gamma and delta zones in highly myopic eyes, concluding that axial elongation primarily drives the stretching and enlargement of these zones. Their findings suggest that axial elongation induces structural changes, such as an enlarged optic disc and optic disc rotation in the ONH. [43]

Importantly, our proposed network independently identified and captured all these structural changes associated with both healthy and diseased conditions, without requiring additional clinical input, underscoring the potential of automated systems in advancing the understanding of ONH morphologies.

High myopia is a significant risk factor for the development and progression of glaucoma, with individuals exhibiting a higher prevalence of glaucoma compared to those without myopia or with mild myopia. [41-47] Structural changes in highly myopic eyes, such as elongated axial length, thinner sclera, and tilted optic discs, increase susceptibility to glaucomatous damage and complicate differentiation from myopia-induced changes. Common glaucoma indicators, such as optic disc cupping and RNFL thinning, often overlap with alterations caused by high myopia, making diagnosis particularly challenging. Additionally, altered biomechanics of the lamina cribrosa and extensive peripapillary atrophy in highly myopic eyes can exacerbate glaucomatous changes. These factors necessitate more intensive monitoring and earlier screening in individuals with high myopia to prevent disease progression and vision loss.

Several structural alterations from HM may predispose myopic eyes to accelerated glaucomatous damage. [44-46] Notably, the HMG ONH may be at greater risk of severe progression due to the pronounced structural changes observed (see Fig. 6e and 6f). A longitudinal observational study found that in patients with myopic glaucoma, higher degrees of myopia were associated with faster rates of visual acuity loss and the development of central visual field defects, indicating a direct correlation between myopia severity and rapid glaucomatous progression. [44] Additionally, Hsueh et al. reported that for each diopter increase in myopia, the risk of developing glaucoma increased by approximately 20%, with more severe myopia linked to faster progression rates, particularly in normal-tension glaucoma. [45] Kim et al. further identified high myopia as a key risk factor for fast visual field progression in glaucoma, attributing this to structural changes such as axial elongation and increased optic nerve susceptibility to damage. [46] These findings underscore the need for tailored management strategies in highly myopic patients to mitigate the risk of rapid glaucoma progression. The proposed study aims to assist clinicians in identifying early structural differences, enabling timely interventions and the development of appropriate management strategies to address the unique challenges posed by high myopia and its associated risks.

This study has several limitations that warrant further discussion. First, glaucoma severity levels were not incorporated into the analysis. Since the structural signature of the ONH varies with glaucoma severity, including this parameter would likely improve the ability to differentiate ONH structures under various pathological conditions more robustly. Second, while OCT captures portions of the anterior and peripapillary sclera, it cannot fully visualize the posterior sclera and deeper scleral structures due to limited penetration depth and signal attenuation. Consequently, only the visible portions of the sclera and LC were segmented in this study. As a result, the reported scleral and LC thickness values may not represent the actual thickness of these layers. Third, the dataset size was relatively small. Although various data augmentation strategies were applied to the point clouds, such as jittering, rigid body rotations along the anterior-posterior axis, and small rigid body translations, the network's performance could likely be further improved with a larger dataset. Fourth, all OCT volumes were acquired using a single device (Spectralis). As scan quality and segmentation accuracy can vary across different devices, [48] it is essential to validate the generated point clouds with images obtained from multiple machines to ensure robustness and generalizability. Finally, the study did not include additional clinical data, such as glaucoma type, demographic parameters, or visual field loss values. Incorporating these variables in future research could enhance the network's performance and provide more comprehensive insights into ONH structural changes associated with glaucoma and myopia.

In summary, this study demonstrated the effective use of point clouds to sparsely represent the complex geometry of the ONH, the application of an autoencoder for point cloud reconstruction and classification, and the use of PCA for dimensionality reduction and visualization. The proposed methodology successfully highlighted the 3D structural phenotypes of the ONH associated with healthy, glaucomatous, and myopic conditions and leveraged these structural signatures for robust classification. This framework holds potential to assist clinicians and researchers in gaining deeper insights into ONH structure and its alterations under pathological conditions.

**Acknowledgements:**

We acknowledge funding from (1) the donors of the National Glaucoma Research, a program of the BrightFocus Foundation, for support of this research (G2021010S [MJAG]); (2) the "Retinal Analytics through Machine learning aiding Physics (RAMP)" project that is supported by the National Research Foundation, Prime Minister's Office, Singapore under its IntraCreate Thematic Grant "Intersection Of Engineering And Health" - NRF2019-THE002-0006 awarded to the Singapore MIT Alliance for Research and Technology (SMART) Centre [MJAG/AT/GB], (3) the National Medical Research Council, Singapore (MOH-000435) [TA], (4) the NMRC-LCG grant 'TAckling & Reducing Glaucoma Blindness with Emerging Technologies (TARGET)', award ID: MOH-OFLCG21jun-0003 [MJAG], (5) the Emory Eye Center (Emory University School of Medicine, Start-up funds, MJAG), (6) a Challenge Grant from Research to Prevent Blindness, Inc. to the Department of Ophthalmology at Emory University, and (7) from the NIH grant P30EY06360 to the Atlanta Vision Community.

**Appendix:**

**Point cloud construction from an OCT volume:**

To create a point cloud from a segmented OCT volume, the anterior and posterior boundaries of the RNFL, ORL, sclera, and LC tissue layers from a bscan were identified using the Canny edge detection filter from the scikit-image library, excluding the GCL+IPL, RPE, and choroid layers (see Fig. 1d). The boundary edges of these layers were converted into a set of 3D boundary points and stored in an array (see Fig. 1e). This process was repeated for all bscans in the OCT volume to construct a 3D point cloud, providing a sparse representation of the complex 3D structure of the ONH. Fig. 1f shows the point cloud generated from the sample OCT volume. The GCL+IPL, RPE, and choroid layers were excluded because their boundaries could be inferred from adjacent layers. For instance, the posterior edge of the RNFL and the anterior edge of the ORL define the boundaries of the GCL+IPL, eliminating the need to extract points specifically from the GCL+IPL edges. This approach reduced the total number of points in the point cloud, allowing for a higher density of points in critical regions. Finally, the point clouds were down sampled to 4096 points. Each point in the cloud was assigned a marker indicating the tissue layer from which it was extracted, enabling layer visualization by assigning unique colors to different layers (see Fig. 1f). The resulting point cloud had a shape of (4096 × 4), where 4096 represents the number of points, and the four dimensions correspond to the x, y, z coordinates and the layer marker of each point.

**Ensemble network for point cloud reconstruction and classification**

Many studies have used autoencoder to extract important features from OCT images as well as from point clouds. Qi et al. proposed a novel deep learning architecture, *PointNet*, designed to process unordered 3D point cloud data for tasks like 3D classification and segmentation.[17][17] PointNet directly operates on point sets, preserving the permutation invariance and unique structure of 3D data. The model is composed of a series of shared multilayer perceptrons (MLP) and a symmetric function (max-pooling) to learn spatial features. Additionally, PointNet uses a spatial transformer network to align and normalize input data, improving model performance by standardizing point cloud orientation. Groueix and co-authors presented a method for generating the surfaces of 3D shapes using a framework called *AtlasNet*, by representing a 3D shape as a collection of parametric surface elements, which allows for a more natural surface representation.[25] The authors used 2D patches to reconstruct 3D point clouds. AtlasNet offered several advantages, including greater precision, better generalization, and the ability to generate shapes at any resolution without running into memory problems.

In this study, we developed a specialized ensembled autoencoder network using the PointNet as the encoder and the AtlasNet as the decoder and a network with dense layers as a classifier (see Fig. 2). The PointNet had feature transform blocks and shared MLP blocks (same as the original work of Qi et al.[17] with a few modifications to the last few layers to obtain the desired latent space dimension) that gradually reduced the input point cloud of shape (4096,3) in to "k" number of latent dimensions. The decoder architecture was similar the AtlasNet as proposed by Groueix et al.[25] but with 32 decoding submodels. Each of these decoding submodels took a parametric grid and the "k" number of latent dimensions to recreate a feature space of dimension (128,3). Subsequently, the output of all 32 decoding submodels were concatenated to reproduce final predicted point cloud of shape (4096,3). The classifier branch utilized a series of MLP layers, followed by a softmax layer for classification. This branch accepted the "k" number of latent dimensions to predict the final classes (i.e., H, HM, G, and HMG) of the point cloud. In the proposed model, the encoder extracted important latent features from the input point cloud, whereas the decoder and classifier used those latent features for point cloud reconstruction and classification, respectively. To the best of our knowledge, no existing literature has extracted feature embeddings from 3D point clouds of the ONH and used them for simultaneous reconstruction and classification.

| Study | Sex (% of female) | Age (mean± SD) | H | HM | G | HMG | Total |
|---|---|---|---|---|---|---|---|
| Cohort 1 | 56 | 61±9.0 | 67 | 91 | 7 | 104 | 269 |
| Cohort 2 | 46 | 66±6.7 | 189 | 3 | 220 | 4 | 416 |
|  |  |  | 256 | 94 | 227 | 108 | 685 |

*Table1: Subject demographics and number of subjects in each category*

| Layer | AUC (micro average) during 5-fold cross validation |
|---|---|
| All layers (RNFL, ORL, Sclera and LC) | 0.92±0.03 |
| RNFL | 0.89±0.02 |
| GCL+IPL and ORL | 0.87±0.04 |
| Choroid | 0.78±0.02 |
| Sclera and LC | 0.85±0.03 |

*Table 2: Diagnostic capability of each layer with micro-average AUC during 5-fold cross-validation*

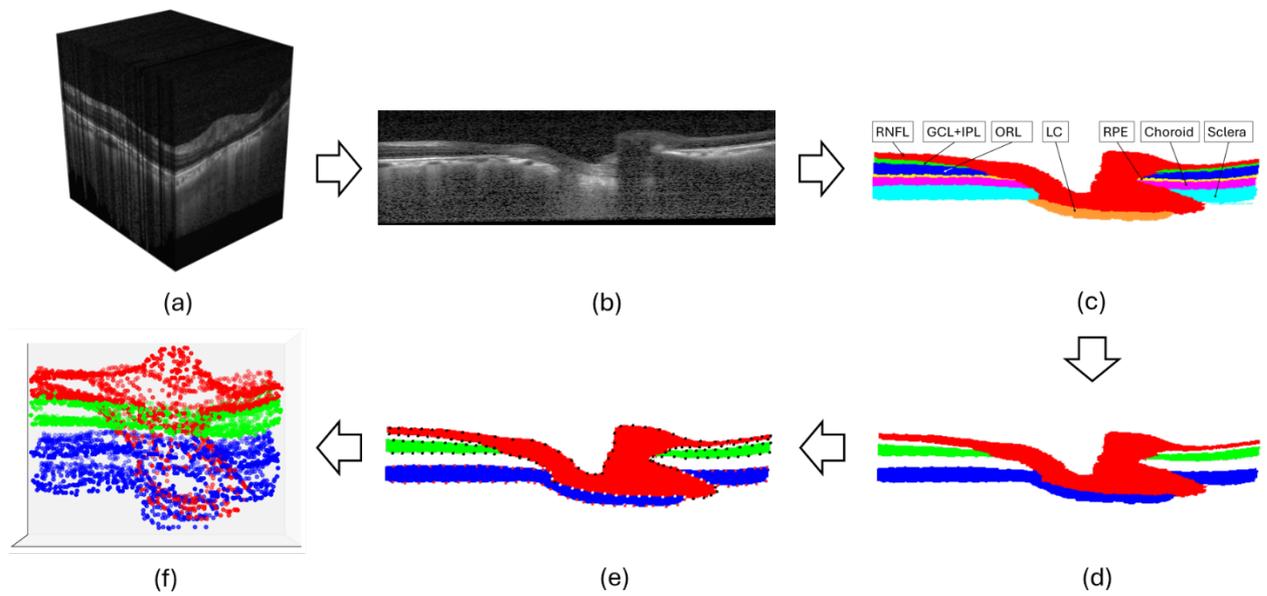

*Figure 1: a) A raw OCT volume, b) A sample BScan from the OCT volume, c) Segmented BScan with seven tissue layers delineated using Reflectivity. Retinal Nerve Fiber Layer (RNFL) in red, Ganglion Cell Layer + Inner Plexiform Layer(GCL+IPL) in green, Other Retinal Layers (ORL) in blue, Retinal Pigment Epithelium (RPE) in yellow, Choroid in pink, Sclera in cyan, and Lamina Cribrosa (LC) in orange. d-e) Boundary point extraction from RNFL, ORL, and sclera + LC layers. F) 3D point cloud generated from the OCT volume*

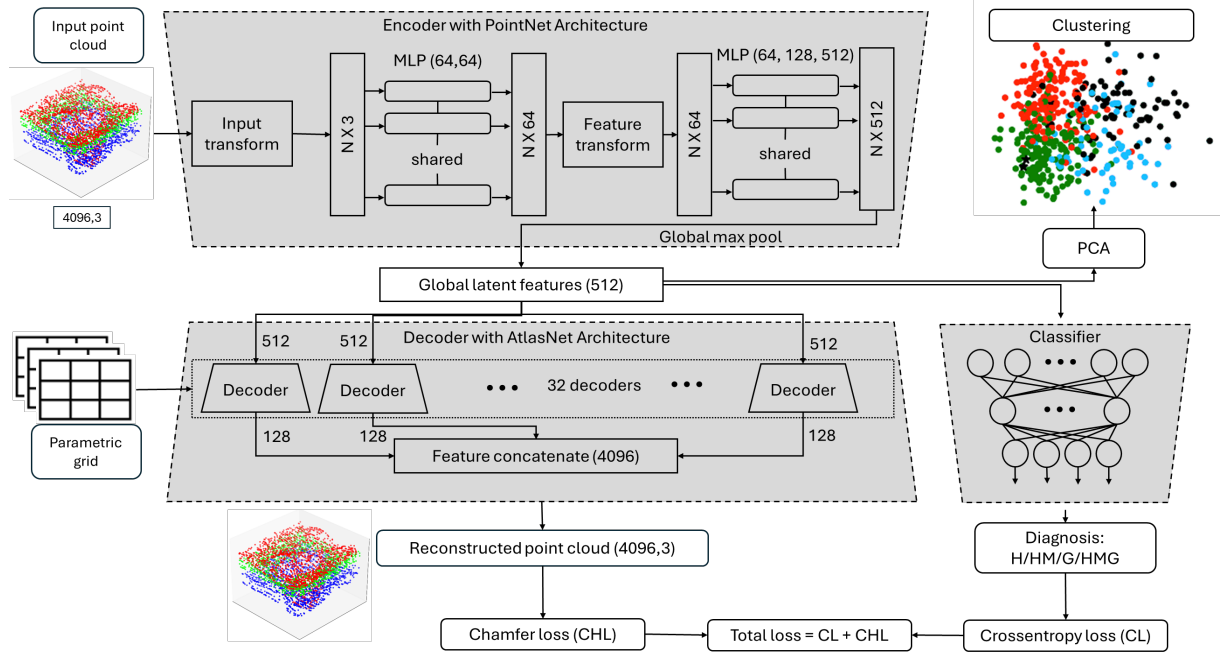

*Figure 2: The proposed ensemble network architecture for simultaneous reconstruction and classification of point clouds. The network had three branches: an encoder, a decoder, and a classifier. A fourth branch that was not part of training process (PCA: Principal Component Analysis) was used for clustering.*

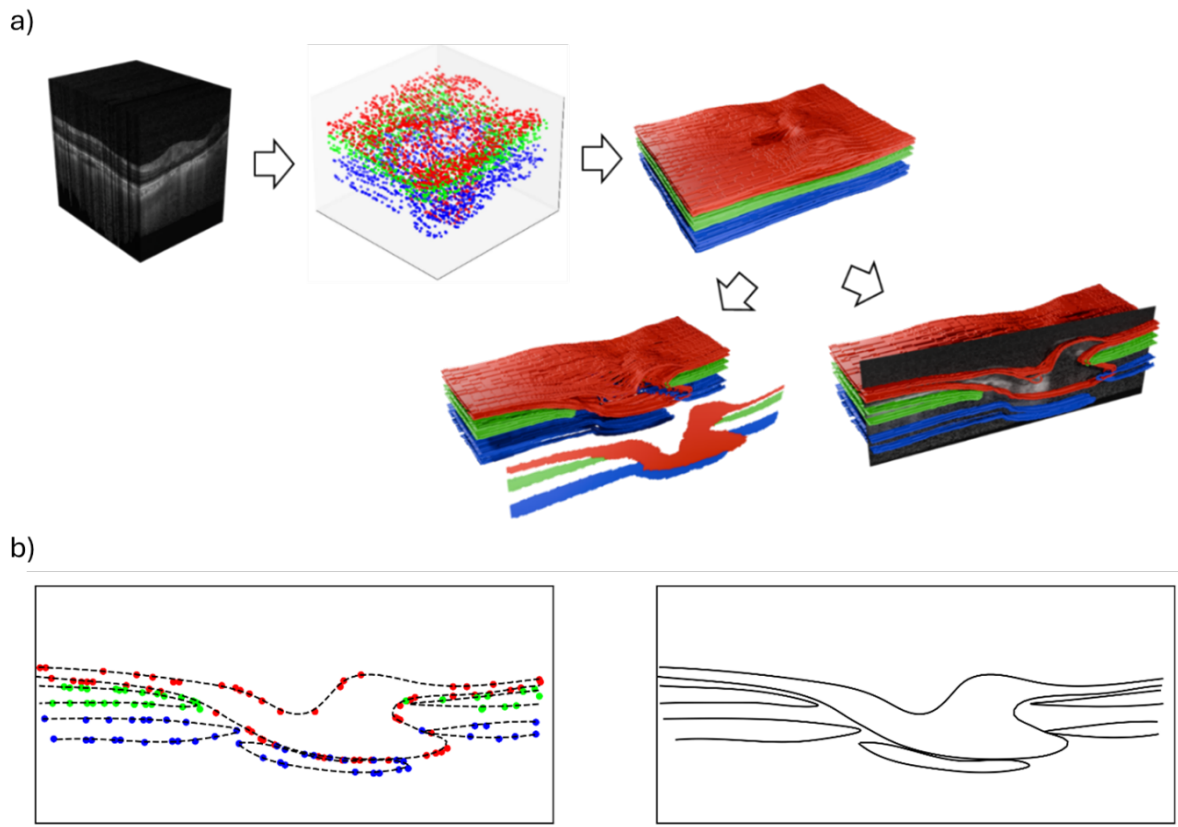

*Figure 3: a) Left to Right: A raw OCT volume and the corresponding point cloud. The point cloud was rendered in Blender (Project Blender, Blender Foundation) for a better visualization. A cut section of the rendered model with the central segmented B-scan (left) and raw B-scan (right) are shown for comparison. Also, the cut section of the rendered model with the central raw B-scan given for comparison. b) The points of the point cloud lying on the central B-scan. Univariate splines are fitted through the points to illustrate the boundary of each layer.*

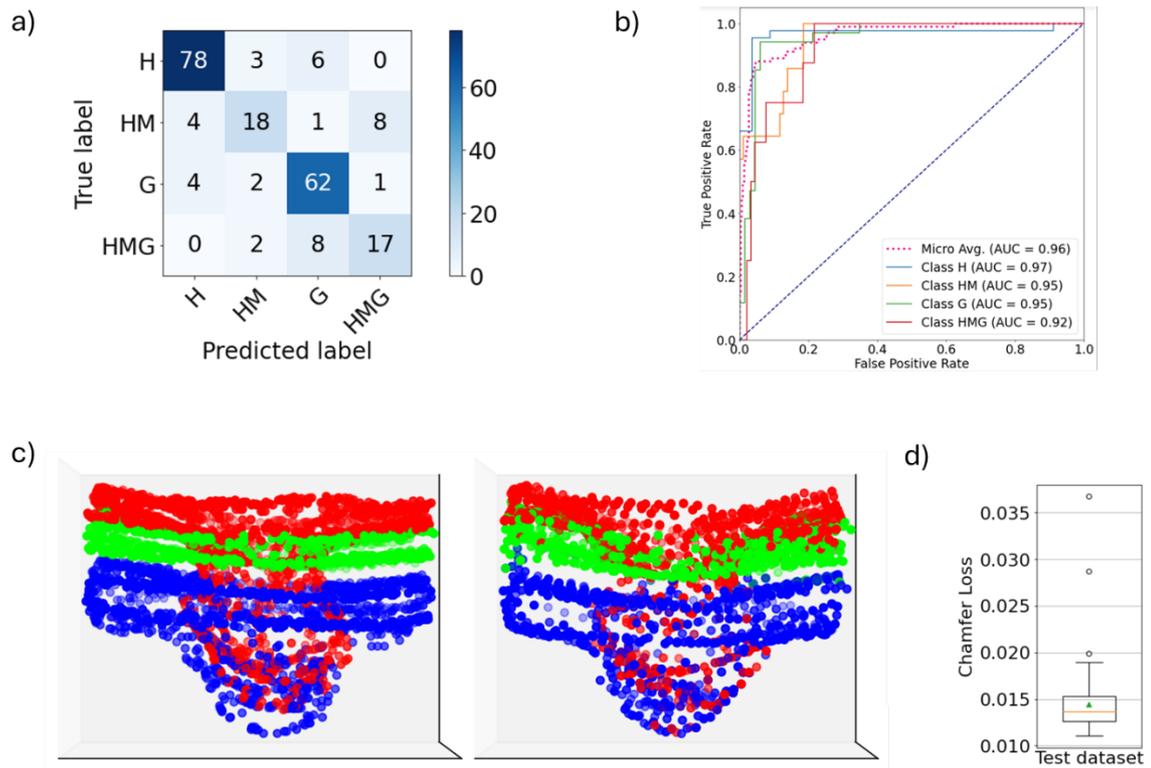

*Figure 4: Classification and Reconstruction network performance on an independent test dataset. a) confusion matrix showing the numbers of correctly classified eyes in each class b) ROC curve (H: Healthy, HM: High myopia, G: Glaucoma, and HMG: High myopia with glaucoma). c) Left: Original point cloud form the test dataset, Right: Reconstructed point cloud by the network. d) Chamfer loss for reconstruction on test dataset*

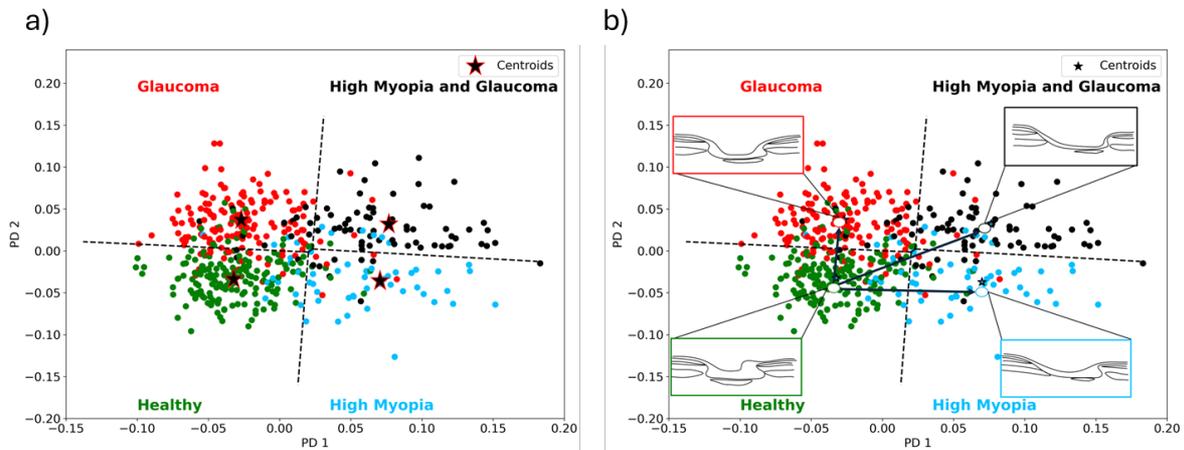

*Figure 5: a) Visualization of 3D point clouds in 2D after reducing the latent vector using PCA. Each dot in the figure represents a point cloud. The green, red, cyan, and black dots correspond H, G, HM, and HMG classes, respectively. The k-means clustering approach successfully identified the boundaries of the clusters (black dashed lines). The centroids of each cluster are highlighted using star markers. b) Structural changes in the ONH while transitioning between clusters. To explore the structural changes, a point from the healthy cluster (large white dot with green edge) was selected and its value was altered incrementally over twenty steps to move it toward a different cluster (black arrows shows the direction of the transition). At each step, the decoder reconstructed a point cloud from the updated latent values. Inset images show the central B-scan of the reconstructed point cloud at the start and end of the transition.*

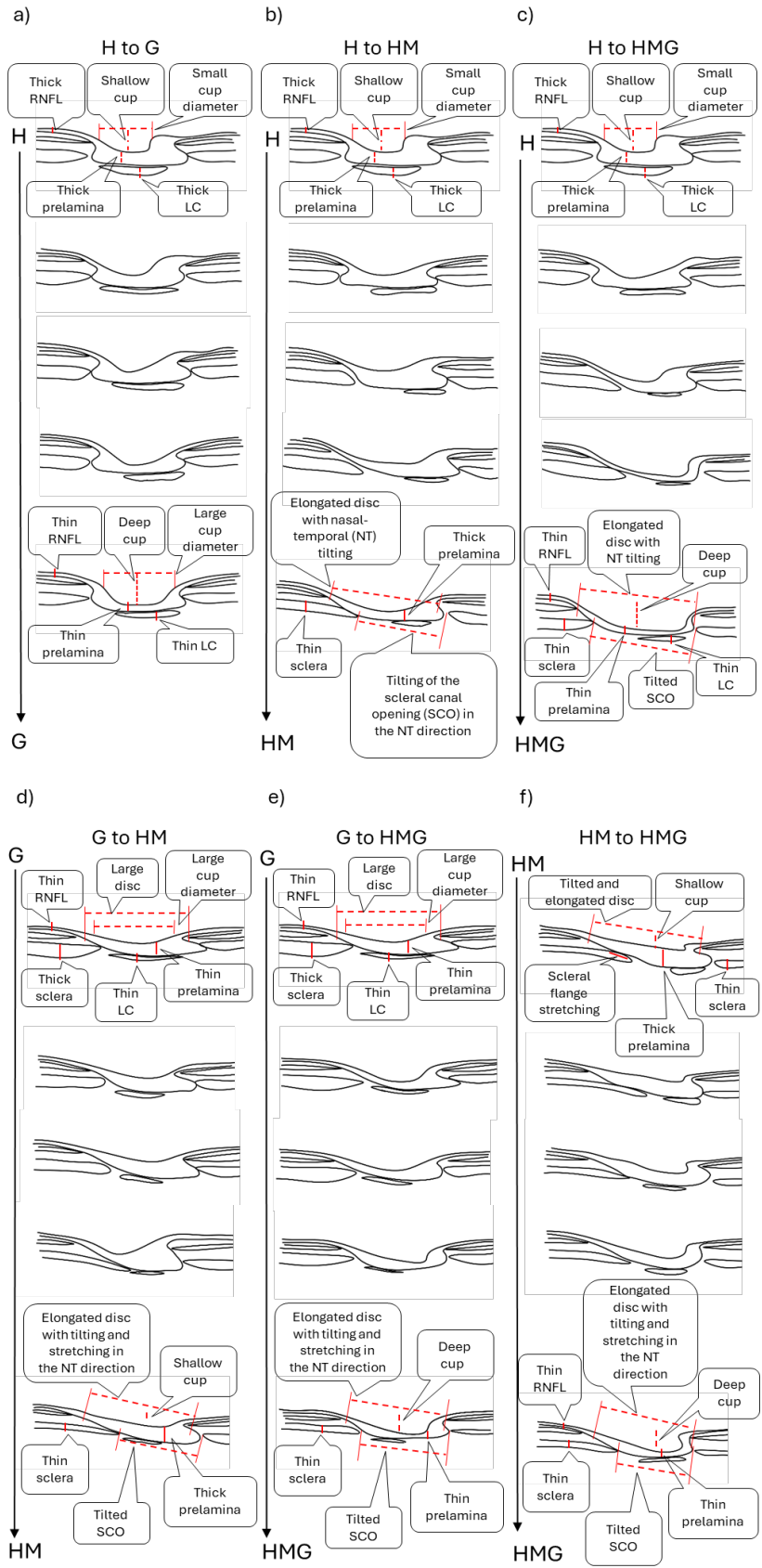

*Figure 6: Distinct structural changes were observed during the morphing of the ONH from one cluster to another. Each column displays a sequence of central B-scans from the point cloud captured at various stages of the morphing process. The top image represents the original*

*baseline B-scan before morphing, while the bottom image shows the final B-scan after morphing. The three images in between depict the morphed B-scans at three intermediate steps.*

Appendix Figure

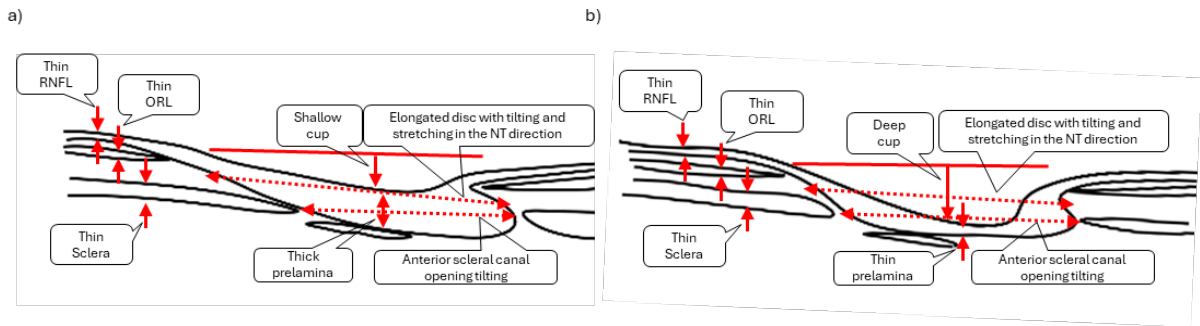

*Figure A1: Distinct structural changes between a) HM and b) HMG ONH.*